\documentclass[journal]{IEEEtran}

\ifCLASSINFOpdf
\else
\fi
\usepackage{algorithm}
\usepackage{algorithmic}
\usepackage{amsmath}
\usepackage{amssymb}
\usepackage{amsfonts}
\usepackage{epsfig}

\newcommand{\beq}{\begin{equation}}
\newcommand{\eeq}{\end{equation}}

\newcommand{\bea}{\begin{eqnarray}}
\newcommand{\eea}{\end{eqnarray}}
\newcommand{\bean}{\begin{eqnarray*}}
\newcommand{\eean}{\end{eqnarray*}}

\newcommand{\bit}{\begin{itemize}}
\newcommand{\eit}{\end{itemize}}

\newcommand{\ben}{\begin{enumerate}}
\newcommand{\een}{\end{enumerate}}

\newcommand {\nbf}[1]{\mbox{\boldmath $#1$} }

\begin{document}
%
\title{Reduced-Complexity SCL Decoding of Multi-CRC-Aided Polar Codes}
%
%
%

\author{Mao-Ching Chiu,~\IEEEmembership{Member,~IEEE,}
\thanks{M.-C. Chiu is with the Department of
Communications Engineering, National Chung Cheng University,
Min-Hsiung, Chia-Yi, 621, Taiwan, R.O.C. (e-mail:
ieemcc@ccu.edu.tw). }
and Wei-De Wu%
\thanks{W.-D. Wu is with the MediaTek, Inc,
No. 1, Dusing 1st Rd., Hsinchu Science Park, Hsinchu City 30078, Taiwan, R.O.C. (e-mail:
weide.wu@mediatek.com). }
}

\maketitle

\begin{abstract}
Cyclic redundancy check (CRC) aided polar codes are capable of achieving better performance than low-density parity-check (LDPC) codes under the successive cancelation list (SCL) decoding scheme. However, the SCL decoding scheme suffers from very high space and time complexities. Especially, the high space complexity is a major concern for adopting polar codes in modern mobile communication standards. In this paper, we propose a novel reduced-complexity  successive cancelation list (R-SCL) decoding scheme which is effective to reduce the space complexity. Simulation results show that, with a $(2048, 1024)$ CRC-aided  polar code, the R-SCL decoders with 25\% reduction of space complexity and 8\% reduction of time complexity can still achieve almost the same performance levels as those decoded by SCL decoders. To further reduce the complexity, we propose a multi-CRC coding scheme for polar codes. Simulation results show that, with a $(16384, 8192)$ multi-CRC-aided polar code, a R-SCL decoder with about 85\% reduction of space complexity and 20\% reduction of time complexity results in a worst performance loss of only 0.04dB.


\end{abstract}


%
\IEEEpeerreviewmaketitle

\section{Introduction}
%
%
%
%
\IEEEPARstart{P}{olar} codes \cite{Arikan2009} are the first family of codes that achieve the capacity of
symmetric binary-input discrete memoryless channels under a low-complexity successive cancelation (SC) decoding algorithm as the code block length $N$ approaches infinity.
They are constructed from the generator matrix $\nbf{G}_2^{\otimes n}$ with $\nbf{G}_2=\left[{1 \atop 1} {0 \atop 1}\right]$, where $\otimes n$ denotes the $n$th Kronecker power. It has been shown in \cite{Arikan2009}, that under perfect successive cancelation (SC) decoding, the channels seen by individual bits start polarizing as $n$ grows large. The bit channels approach either a noiseless channel or a pure-noise channel. The fraction of noiseless channels is close to the channel capacity. Therefore, the noiseless channels, termed unfrozen bit channels, are selected for transmitting message bits while the other channels, termed frozen bit channels, are set to fixed values known by both encoder and decoder.

Though polar codes asymptotically achieve the channel capacity, the performance of polar codes at short to moderate block lengths is disappointing under the SC decoding algorithm. There are two reasons as observed in \cite{Tal2011, Tal2012}. The first reason is that the performance of the SC decoder is significantly degraded compared to that of the maximum-likelihood (ML) decoder. The second reason is that polar codes are inherently weak at short to moderate block lengths. To improve the performance of the decoder, a successive cancelation list (SCL) decoding algorithm was proposed \cite{Tal2011, Tal2012} which performs almost the same as the ML decoding scheme as the list size $L$ is large. However, since polar codes are weak, the performance levels of polar codes even under the ML decoding scheme are inferior to those of low-density parity-check (LDPC) codes. To strengthen polar codes, a concatenation scheme of cyclic redundancy check (CRC) codes and polar codes, termed CRC-aided polar codes,  was found to be effective to improve the performance \cite{Tal2011, Tal2012} under the SCL decoding scheme. It has been shown that CRC-aided polar codes under SCL decoding are capable of achieving better performance levels than those of LDPC codes and turbo codes \cite{Tal2011, Tal2012, Li2012}.

The space and time complexities of the SCL decoder with list size $L$ are $O(LN)$ and $O(LN\log_2 N)$, respectively, where $N$ is the block length. As a result, the complexities of the SCL decoder become very huge as $L$ or $N$ is large. Recently, the family of CRC-aided polar codes has been considered as a candidate for the fifth generation (5G) mobile communication standard \cite{Tdoc-84b-2016-5G-coding-candidates}. 
However, the high space complexity of the SCL decoder becomes a major concern by adopting CRC-aided polar codes in the 5G mobile communication standard \cite{R1-164360-2016}. Therefore, it is very critical to reduce the space complexity of the SCL decoder. There are several tree-pruning techniques proposed in \cite{Chen2013, Chen2015} to reduce the {\em averaged space and time complexities} of the SCL decoder and hence to reduce the energy consumption of the SCL decoder. In tree-pruning techniques, the candidate paths of small reliability values are eliminated from the list during SCL decoding steps. This method can reduce the computational overhead to extend the paths which are not likely to be the correct codeword. However, the worst-case space and time complexities remain $O(LN)$ and $O(LN\log_2N)$, respectively. In \cite{Li2012}, it was observed that, for most of the cases, the SCL decoder with very small $L$ can successfully decode the information bits, and there are very few cases that need very large $L$ for successful decoding. Therefore, in order to reduce the
decoding complexity, an adaptive SCL decoder for CRC-aided polar codes was proposed \cite{Li2012}. The adaptive SCL decoder initially uses a very small $L$ for decoding. If there is no candidate path passing the CRC check, the decoder iteratively increases $L$, until $L$ reaches a predefined value $L_{\max}$. It is obvious that adaptive SCL decoder can only improve the {\em averaged space and time complexities}. The worst-case space and time complexities remain $O(L_{\max}N)$ and $O(L_{\max}N\log_2N)$, respectively.

In this paper, we propose a reduced-complexity SCL (R-SCL) decoder for polar codes, targeting to reduce the worst-case complexity. For SCL decoder with list size $L$, $L$ memory blocks with each block size of $2^{n-m}$ are required to store the log-likelihood ratios (LLRs) at the $m$th intermediate stage for $m=1, \ldots, n$ in the graph representation of polar codes \cite{Tal2011, Tal2012, Balatsoukas-Stimming2015}. Consequently, the space complexity of the SCL decoder is given by $O(L\sum_{m=1}^n 2^{n-m}) = O(LN)$. In the R-SCL decoder, we propose to apply smaller or equal number of memory blocks for the stage with larger memory block size. Therefore, the numbers of memory blocks are now specified by a vector $\nbf{L} = [L_1, L_2, \ldots, L_n]$ where $L_m \le L_m'$ for $m \le m'$. In this way, the space and time complexities can be effectively reduced. Simulation results will verify that the R-SCL decoder with 25\% reduction of space complexity and 9\% reduction of time complexity can achieve little performance loss w.r.t. the SCL decoder.

To further reduce the space and time complexities of the R-SCL decoder, we propose a novel multi-CRC coding scheme. The idea of multi-CRC-aided polar codes was first proposed in \cite{Guo2016}. The method proposed in \cite{Guo2016} divides the entire $K$-bit message block into $M$ sub-blocks. All sub-blocks have equal length of $K'=K/M$, and $r$ CRC bits are appended to each sub-block. In order to reduce the decoding delay, the modified SCL decoder proposed in \cite{Guo2016} outputs each sub-block as early as possible during the SCL decoding. When the decoding level reaches the last bit of a sub-block, the decoder applies CRC detection on the sub-block immediately among all candidate paths in the list. The most reliable path that passes the CRC check is selected as the estimation of the information sub-block. In this way, the latency can be reduced and the memory to store the {\it hard-decision} bits of all paths can be released. However, the space complexity of the modified SCL decoder remains the same, because only memory to store the  {\it hard-decision} bits is released  \cite{Guo2016}. The proposed multi-CRC coding scheme is different from that proposed in \cite{Guo2016}. In the new multi-CRC coding scheme, the $N=2^n$ bit channels are partitioned into $M = 2^s$ sub-blocks, and each sub-block contains $2^{n-s}$ bit channels and its own CRC bits. 
The CRC bits for each sub-block can help the R-SCL decoder to further reduce the number of memory blocks at stages smaller than or equal to $s$ by only retaining fewer CRC-passed candidate paths at decoding levels of $(j+1) 2^{n-s}-1$ for $j = 0, \ldots, M-1$. Simulation results show that, with the multi-CRC-aided polar codes, the R-SCL decoders can achieve 85\% reduction of space complexity and 20\% reduction of time complexity at performance penalty no larger than $0.04$ dB w.r.t. the conventional SCL decoder with single-CRC-aided polar codes.

This paper is organized as follows. Section \ref{sec:back} gives a background introduction of polar codes. The SC decoder and SCL decoder are briefly reviewed in Section \ref{sec:SC} and Section \ref{sec:SCL}, respectively. Section \ref{sec:R-SCL} proposes the reduced-complexity SCL (R-SCL) decoder. The multi-CRC-aided polar code with R-SCL decoder is proposed in Section \ref{sec:mCRC}. Simulation results are given in Section \ref{sec:simu}. Conclusions are drawn in Section \ref{sec:conc}.

Notations: Throughout this paper, matrices and vectors are set in boldface, with uppercase
letters for matrices and lower case letters for vectors. An $n$-tuple vector $\nbf{x}$ is denoted as $\nbf{x} = [x_0, x_1, \ldots, x_{n-1}]$ with the indices starting from 0 (instead of 1 for normal vector representations). The notation $\nbf{x}_{a}^{b}$ means the sub-vector $[x_a, x_{a+1}, \ldots, x_{b}]$ if $b \geq a$ and null vector otherwise. Set
quantities such as ${\cal A}$ are denoted using the calligraphic font, and the cardinality of the set ${\cal A}$ is
denoted as $|{\cal A}|$. The binary representation of an integer $i$ with $0\leq i <2^n$ is denoted by an $n$-tuple binary vector $\nbf{i} = [i_0, i_1, \ldots, i_{n-1}]$ satisfying $i = \sum_{j=0}^{n-1} i_j 2^{j}$.

\section{Background}
\label{sec:back}
\begin{figure}[!t]
\centering
\includegraphics[width=0.8\columnwidth]{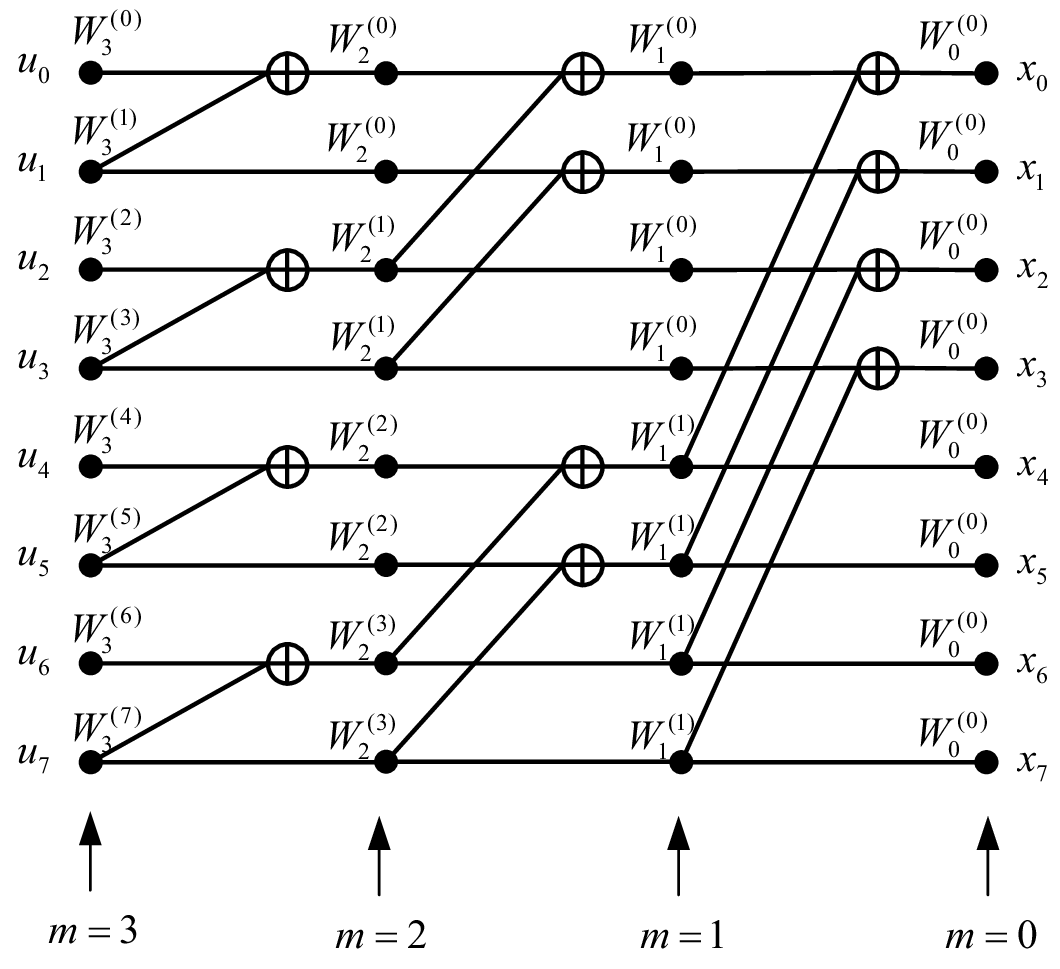}
\caption{Graph representation of a polar code of block length $N=8$.}
\label{fig:enc8}
\end{figure}

Polar codes are  constructed from the generator matrix $\nbf{G}_2^{\otimes n}$ with $\nbf{G}_2=\left[{1 \atop 1} {0 \atop 1}\right]$, where $\otimes n$ denotes the $n$th Kronecker power.
A codeword of a polar code of length $N = 2^n$ without bit-reversal matrix can be represented by
\beq
      \nbf{x} = \nbf{u} \nbf{G}_2^{\otimes n}, \label{eq:genm}
\eeq
where $\nbf{u} = [u_0, u_1, \ldots, u_{2^n-1}]$ is the message bits and $\nbf{x} = [x_0, x_1, \ldots, x_{2^n-1}]$ is the codeword bits. In this paper, we employ polar codes without bit-reversal matrix and the graph representation is different from that proposed in \cite{Arikan2009}. A polar code of block length $2^n$ can be represented by a graph with $n$ sections of trellises as given in \cite{Arikan2009} which is called the standard graph of the polar code. It has been indicated in \cite{Hussami2009a} that for a polar code of block length $2^n$, there exist $n!$ different
graphs obtained by different permutations of the $n$ layers of trellis connections. We consider to represent a polar code with reverse ordering of its standard graph. Figure \ref{fig:enc8} shows an example graph for $N = 8$ with reverse ordering of the standard graph.

The codeword $\nbf{x}$ obtained from (\ref{eq:genm}) is than transmitted via $N$ independent uses of the binary input discrete memoryless channel (B-DMC) $W:{\cal X} \rightarrow {\cal Y}$, where ${\cal X} = \{0, 1\}$ denotes the input alphabet, ${\cal Y}$ denotes the output alphabet, and $W(y|x)$ denotes the channel transition probabilities. The conditional distribution of the output ${\nbf Y} = {\nbf{y}}$ given the input ${\nbf{X}} = \nbf{x}$, denoted as $W^N(\nbf{y}|\nbf{x})$, is given by
\[
W^N(\nbf{y}|\nbf{x}) = \prod_{i=0}^{N-1} W(y_i|x_i).
\]
The distribution of $\nbf{Y}$ conditioned on $\nbf{U}=\nbf{u}$, denoted as $W_n(\nbf{y}|\nbf{u})$, is given by
\[
     W_n(\nbf{y}|\nbf{u}) = W^N(\nbf{y}|\nbf{u}\nbf{G}_2^{\otimes n}).
\]
The polar code of length $N=2^n$ transfers the original $2^n$ identical channels $W$ into $2^n$ synthesized channels, denoted as $W_n^{(i)}: {\cal X} \rightarrow {\cal Y}^N \times {\cal X}^i$ for $i \in {0, \ldots, 2^n-1}$ with the transition probability given by
\[
    W_n^{(i)}(\nbf{y}, \nbf{u}_0^{i-1}|u_i) \equiv \sum_{\nbf{u}_{i+1}^{N-1} \in {\cal X}^{N-i-1}} \frac{1}{2^{N-1}}W_n(\nbf{y}|\nbf{u}).
\]
It has been shown in \cite{Arikan2009} that as $n$ grows large, the synthesized channels start polarizing. They approach either a noiseless channel or a pure-noise channel. The fraction of noiseless channels is close to the channel capacity. Therefore, the noiseless channels are selected for transmitting message bits while the other channels are set to fixed values known by both encoder and decoder. In the code design, a polar code of dimension $K$ is generated by selecting the $K$ least noisy channels among $W_n^{(i)}$ and denotes the indices of the $K$ least noisy channels as a set ${\cal A}$. Define $\nbf{u}_{\cal A}$ as a sub-vector of $\nbf{u}$ formed by the elements of $\nbf{u}$ with indices in ${\cal A}$. Only the sub-vector $\nbf{u}_A$, termed {\em unfrozen bits}, is employed to transmit message bits. The other bits $\nbf{u}_{{\cal A}^c}$, termed {\em frozen bits}, are set to fixed values known by both encoder and decoder. In this paper, we set the frozen bits to all zeros.

\section{Successive Cancelation (SC) Decoding of Polar Codes}
\label{sec:SC}
The SC decoder of polar codes was proposed in \cite{Arikan2009}. In this section, for future reference, we give a brief review of SC decoder using log-likelihood ratios (LLR).

The SC decoder in probability domain is based on the calculation of a pair of conditional probabilities, i.e., $W_n^{(i)}(\nbf{y}, \hat{\nbf{u}}_{0}^{i-1}|u_i)$ for $u_i \in \{0, 1\}$, for the decision of the $i$th bit $u_i$, where $\hat{\nbf{u}}_{0}^{i-1}$ are the decisions on previous decoding steps. Let $\bar{N} = 2^m$ with $0 \leq m\leq n$. It has been shown in \cite{Arikan2009} that the probability pairs can be calculated using the recursions
\bean
    &&W_m^{(2i)}(\nbf{y}_0^{\bar{N}-1}, \nbf{u}_0^{2i-1}|u_{2i}) \\
    &&= \sum_{u_{2i+1}} \frac{1}{2}W_{m-1}^{(i)}(\nbf{y}_{0, e}^{\bar{N}-1}, \nbf{u}_{0, e}^{2i-1} \oplus \nbf{u}_{0, o}^{2i-1}|u_{2i} \oplus u_{2i+1}) \\
    &&\ \ \ \ \ \ \ \ \ \ \cdot W_{m-1}^{(i)}(\nbf{y}_{0, o}^{\bar{N}-1}, \nbf{u}_{0, o}^{2i-1}|u_{2i+1}),
\eean
and
\bean
    &&W_m^{(2i+1)}(\nbf{y}_0^{\bar{N}-1}, \nbf{u}_0^{2i}|u_{2i+1}) \\
    &&= \frac{1}{2}W_{m-1}^{(i)}(\nbf{y}_{0, e}^{\bar{N}-1}, \nbf{u}_{0, e}^{2i-1} \oplus \nbf{u}_{0, o}^{2i-1}|u_{2i} \oplus u_{2i+1}) \\
    &&\ \ \ \ \cdot W_{m-1}^{(i)}(\nbf{y}_{0, o}^{\bar{N}-1}, \nbf{u}_{0, o}^{2i-1}|u_{2i+1}).
\eean
Figure \ref{fig:enc8} shows the relative positions of $W_m^{(i)}$ in the graph representation of a polar code of length $N=8$. The index $m$ denotes the $m$th stage of the graph for $m=0,1, \ldots, n$ from right to left. 

For efficient implementations of the decoding operation, some values $W_m^{(i)}$ used to decode current information bit are store in the memory which can be shared to decode the subsequent information bits.
As proposed in \cite{Arikan2009}, a total of $2^{n-m}$ probability pairs is required to store in the memory at stage $m$.
Therefore, the space and time complexities of the SC decoder are $O(N)$ and $O(N\log_2N)$, respectively. However, the probability pairs are prone to underflows, i.e., they approach zero as $n$ is large. Therefore, certain normalization methods are required to normalize the probability pairs in the intermediate stages of the graph. The LLR based SC decoder does not require any normalization and has the advantage that it is numerically stable. In addition, only one value is required to compute which enables area-efficient implementations.

\begin{figure}[!t]
\centering
\includegraphics[width=\columnwidth]{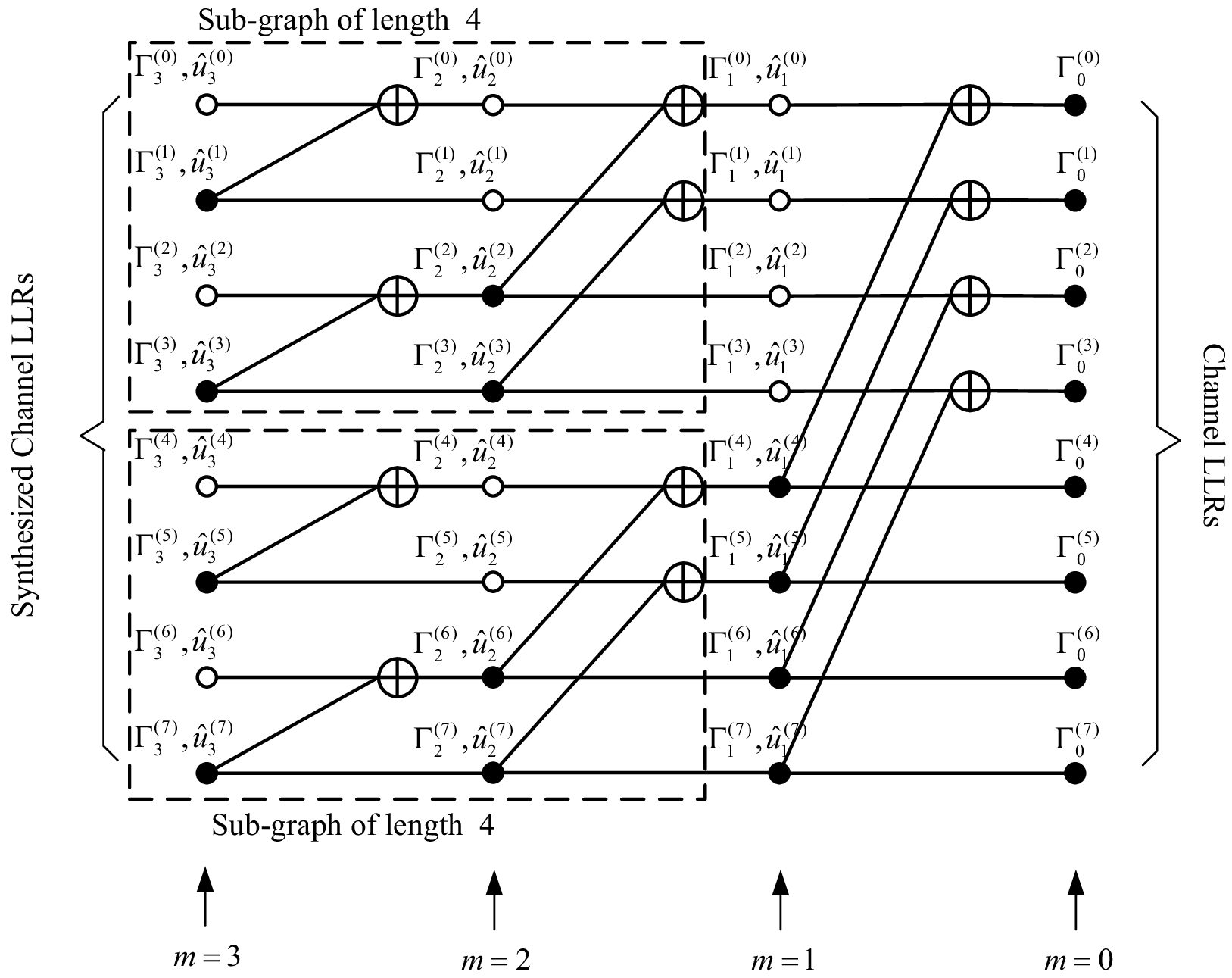}
\caption{SC decoding of a polar code of block length $N=8$.}
\label{fig:dec8}
\end{figure}

Instead to compute $W_n^{(i)}(\nbf{y}, \hat{\nbf{u}}_{0}^{i-1}|u_i)$ for $u_i \in \{0, 1\}$, it is sufficient to compute the LLRs
\beq
     \Gamma_n^{(i)} = \ln\left(\frac{W_n^{(i)}(\nbf{y}, \hat{\nbf{u}}_{0}^{i-1}|0)}{W_n^{(i)}(\nbf{y}, \hat{\nbf{u}}_{0}^{i-1}|1)}\right).
     \label{eq:lni}
\eeq
The decision $\hat{u}_i$ is given by $\hat{u}_i = 0$ for $i \in {\cal A}^c$ and
\[
     \hat{u}_i = \left\{
     \begin{array}{ll}
     0, & \mbox{if $\Gamma_n^{(i)} > 0$} \\
     1, & \mbox{otherwise}
     \end{array}
     \right.,
\]
for $i \in {\cal A}$. Define intermediate LLRs at stage $m$ as $L_m^{(i)}$ for $i=0, \ldots, N-1$ as illustrated in Figure \ref{fig:dec8}.  The LLRs (\ref{eq:lni}) can be computed using the recursions
\bea
   \Gamma_m^{(i)} &=& \Gamma_{m-1}^{(i)} \boxplus \Gamma_{m-1}^{(i+2^{n-m})}, \nonumber \\
   \Gamma_m^{(i+2^{n-m})} &=& (-1)^{\hat{u}_m^{(i)}} \Gamma_{m-1}^{(i)} + \Gamma_{m-1}^{(i+2^{n-m})}, \nonumber
\eea
for $i=0, \ldots, N-1$ and $m = n, n-1, \ldots, 1$, where the binary operator $\boxplus$ is defined by
\[
a \boxplus b \equiv \ln\left(\frac{e^{a+b}+1}{e^a+e^b}\right).
\]
The recursions terminate at $m=0$ with $\Gamma_0^{(i)}$ given as the channel LLRs:
\[
     \Gamma_0^{(i)} = \ln\left(\frac{W(y_i|0)}{W(y_i|1)}\right).
\]
The {\em partial sum} $\hat{u}_m^{(i)}$ are computed starting from $\hat{u}_n^{(i)} = \hat{u}_i$ with
\bean
     \hat{u}_{m-1}^{(i)} &=& \hat{u}_{m}^{(i)} + \hat{u}_{m}^{(i+2^{n-m})},\\
     \hat{u}_{m-1}^{(i+2^{n-m})} &=& \hat{u}_{m}^{(i+2^{n-m})},
\eean
for $m = n, n-1, \ldots, 1$.

Similarly, partial LLRs calculated at each nodes for the $i$th bit can be shared to calculate the LLRs of the $j$th bits for $j > i$. To share the LLRs, some LLRs are required to be stored at each stage. The size of memory to store the LLRs at stage $m$ is $2^{n-m}$ \cite{Arikan2009}. Therefore, the total memory size to store the LLRs is $\sum_{m=1}^n2^{n-m}=2^n-1\approx N$ excluding the memory to store the channel LLRs.

The decoder can be viewed as that implemented using divide-and-conquer procedures. The original polar code of length $2^n$ is divided into two polar codes of length $2^{n-1}$. For example, Figure \ref{fig:dec8} shows a polar code of length $8$ is divided into two polar codes of length 4, viewing as the left part of Figure \ref{fig:dec8} form $m=1$. One is represented as the upper sub-graph with LLR inputs $\Gamma_1^{(0)}, \Gamma_1^{(1)}, \Gamma_1^{(2)}, \Gamma_1^{(3)}$, and the other is represented as the lower sub-graph with LLR inputs $\Gamma_1^{(4)}, \Gamma_1^{(5)}, \Gamma_1^{(6)}, \Gamma_1^{(7)}$. The two polar codes of length $2^{n-1}$ can be further divided into 4 polar codes of length $2^{n-2}$, and so on. Therefore, in general, at stage $m$, we may have $2^{m}$ polar codes of length $2^{n-m}$. 

To decode the $i$th bit, the LLR inputs of each sub-graph at stage $m$ which the $i$th message bit belongs to are calculated and stored in the memory. For example, for $i=1$, the input LLRs, $\Gamma_2^{(0)}$ and $\Gamma_2^{(1)}$, for the upper most sub-graph at stage $m=2$ and the input LLRs, $\Gamma_1^{(0)}$, $\Gamma_1^{(1)}$, $\Gamma_1^{(2)}$, and $\Gamma_1^{(3)}$, for the upper most sub-graph at stage $m=1$ are calculated and store in the memory. Then for $i=2$, only the LLRs at stage $m=2$ are updated using the partial sums $\hat{u_2}^{(0)}$ and $\hat{u_2}^{(1)}$ which is given by
\bean
      \Gamma_2^{(2)} &=& (-1)^{\hat{u}_2^{(0)}} \Gamma_1^{(0)} + \Gamma_1^{(2)},\\
      \Gamma_2^{(3)} &=& (-1)^{\hat{u}_2^{(1)}} \Gamma_1^{(1)} + \Gamma_1^{(3)}.
\eean
Finally, $\Gamma_3^{(2)}$ is calculated by $\Gamma_3^{(2)} = \Gamma_2^{(2)} \boxplus \Gamma_2^{(3)}$ and the estimate $\hat{u}_2$ is decided based on $\Gamma_3^{(2)}$. Then, for $i=3$, the LLR $\Gamma_3^{(3)}$ is calculated by
\[
     \Gamma_3^{(3)} = (-1)^{\hat{u}_3^{(2)}} \Gamma_2^{(2)} + \Gamma_2^{(3)},
\]
and the estimate $\hat{u}_3$ is decided based on $\Gamma_3^{(3)}$.
The procedure is repeated until all the message bits are decoded. Therefore, the entire set of $N\log_2N$ LLRs $\Gamma_m^{(i)}$ can be computed using $O(N\log_2N)$ updates. The memory size to store the LLRs is $O(N)$.

%
%


\section{Successive Cancelation List (SCL) Decoding of Polar Codes}
\label{sec:SCL}
The SC decoding has the drawback that if a bit is not correctly detected, it is not possible to correct it in future decoding steps. Therefore, to improve the performance of the SC decoder, the successive cancelation list (SCL) decoder was proposed in \cite{Tal2011, Tal2012} based on the probability domain. Later, it was shown that the SCL decoder can be implemented based on the LLR domain \cite{Balatsoukas-Stimming2014, Balatsoukas-Stimming2015}. For future reference, we give a brief review of the SCL decoder based on the LLR domain.

The SCL decoding algorithm is based on the binary tree search over the information bits under a complexity constraint that the size of the list is at most $L$. Any information bit sequence can be represented as a path over the binary tree. The SCL decoder searches over the binary tree level-by-level. A each level $i \in {\cal A}$, the decoder extends every candidate path in the list along two paths of the binary tree by appending a bit 0 or a bit 1 to each of the candidate path. Therefore at every $i \in {\cal A}$, the decoder doubles the number of paths up to a predetermined limit $L$. When the number of paths exceeds $L$, only $L$ most reliable paths are retained for further processing at the next level. This procedure is repeated until $i=N-1$. Then, the path of the largest reliability metric is selected as the decoder output. The SCL decoder degenerates to SC decoder when $L=1$.

Let $\hat{\nbf{u}}_0^{i-1}[\ell]$ be one of the paths in the list at level $i-1$, where $\ell \in \{0, \ldots, L-1\}$ denotes the path index in the list. The path metrics of the extended paths $[\hat{\nbf{u}}_0^{i-1}[\ell], 0]$ and $[\hat{\nbf{u}}_0^{i-1}[\ell], 1]$ proposed in \cite{Tal2011, Tal2012} is calculated based on the probability domain using the likelihoods $W_n^{(i)}(\nbf{y}, \hat{\nbf{u}}_0^{i-1}[\ell]|0)$ and $W_n^{(i)}(\nbf{y}, \hat{\nbf{u}}_0^{i-1}[\ell]|1)$.
Therefore, two values have to be computed at each intermediate update and stored at each stage of the graph. Besides, the likelihood values in the probability domain are prone to underflows. To avoid underflow, the likelihood values at each intermediate update are scaled by a common factor \cite{Tal2011, Tal2012} such that the final path metric is proportional to $W_n^{(i)}(\nbf{y}, \hat{\nbf{u}}_0^{i-1}[\ell]|u_i)$ for $u_i \in \{0, 1\}$.

Another implementation of the SCL decoder based on the LLR domain was proposed in  \cite{Balatsoukas-Stimming2014, Balatsoukas-Stimming2015}. It has been proved in \cite{Balatsoukas-Stimming2014, Balatsoukas-Stimming2015} that the path metric defined by
\[
     M_{\ell}^{(i)} = \sum_{j = 0}^ i \ln \left[1+ \exp\left(-(-1)^{\hat{u}_j[\ell]} \Gamma_n^{(j)}[\ell]\right)\right]
\]
is proportional to $\ln \left(W_n^{(i)}(\nbf{y}, \hat{\nbf{u}}_0^{i-1}[\ell]|\hat{u}_i[\ell])\right)$, where
\[
\Gamma_n^{(i)}[\ell] = \ln\left(\frac{W_n^{(i)}(\nbf{y}, \hat{\nbf{u}}_0^{i-1}[\ell]|0)}{W_n^{(i)}(\nbf{y}, \hat{\nbf{u}}_0^{i-1}[\ell]|1)}\right),
\]
which is the LLR of bit $u_i$ given the channel output $\nbf{y}$ and the path $\hat{\nbf{u}}_0^{i-1}[\ell]$. Using the new path metric, the SCL decoder can be implemented using $L$ parallel low-complexity LLR-based SC decoders given in Section \ref{sec:SC} as the building blocks. The metric for the new extended path $[\hat{\nbf{u}}_{0}^{i-1}[\ell], \hat{u}_i[\ell]]$ can be updated with
\[
    M_{\ell}^{(i)} = M_{\ell}^{(i-1)} + \ln \left[1+ \exp\left(-(-1)^{\hat{u}_i[\ell]} \Gamma_n^{(i)}[\ell]\right)\right],
\]
where $M_{\ell}^{(i-1)}$ is the path metric of $\hat{\nbf{u}}_{0}^{i-1}[\ell]$.

During SCL decoding, a decoding path is extended into two candidates, and hence the contents of the LLR values stored in the intermediate stages have to be duplicated with one copy given to the first candidate and the other to the second. Therefore, the copying operation alone would take time $O(LN^2)$ for a naive implementation of the SCL decoder. A cleaver implementation using a so call ``lazy-copy'' technique was proposed in \cite{Tal2011, Tal2012} which enables the memory sharing structure among candidate paths. Recall that the size to store $\Gamma_m^{(i)}$ at stage $m$ is $2^{n-m}$ for the SC decoder. For the SCL decoder, $L$ memory blocks with each block size of $2^{n-m}$ are maintained at stage $m$ to store $\Gamma_m^{(i)}$ and hence the space complexity of the SCL decoder is $O(L\sum_{m=1}^n 2^{n-m}) = O(LN)$.  The memory blocks of stage $m$ are indexed by an integer from $0$ to $L-1$. The ``lazy copy'' technique is to copy the memory indices rather than the contents of the memory. In this way, the time complexity is reduced to $O(LN\log_2N)$ instead of $O(LN^2)$.

As observed in \cite{Tal2011, Tal2012}, even with the ML decoder, the performance of the polar code is still inferior to  that of the LDPC code for a block length of $N=2048$.
This means that with high probability the transmitted codeword is not the ML codeword. However, it was observed that with high probability the transmitted codeword is in the list of the SCL decoder at the last decoding level. Therefore, a simple cyclic redundancy check (CRC) encoding scheme \cite{Tal2011, Tal2012} is applied for a information block of length $K$ by adding $r$ redundant bits. The CRC-aided polar code requires $K+r$ bit channels, i.e., $|{\cal A}|=K+r$, for transmission instead of $K$ bit channels. The CRC-$r$ encoded block can help the decoder to decide which codeword in the list is the transmitted codeword by selecting the CRC-passed candidate of the largest metric among the list at the last decoding level. An example of a $(2048, 1024)$ polar code with CRC-16 under SCL decoding with $L=32$ was found to be capable of achieving almost the same performance as that of the LDPC code of the same length and code rate \cite{Tal2011, Tal2012}. Therefore, CRC-aided polar code is a competitive candidate in future wireless communication standards.

\section{Reduced-Complexity SCL Decoding of Polar Codes}
\label{sec:R-SCL}

Recently, the family of CRC-aided polar codes has been considered as a candidate for the fifth generation (5G) mobile communication standard \cite{Tdoc-84b-2016-5G-coding-candidates}. However, the high space complexity of the SCL decoder is a major concern by adopting polar codes \cite{R1-164360-2016}.
In this paper, we propose a hardware friendly reduced-complexity SCL (R-SCL) decoding algorithm which is suitable for hardware implementations with limited hardware resources.

To investigate the space complexity of the SCL decoder, it is important to determine how many memory blocks are used at each stage when the decoding level $i$ ranges from $0$ to $N-1$. Let ${\cal U}^{(i)}$ be the set of candidates in the list at decoding level $i$, i.e., ${\cal U}^{(i)} = \{\hat{u}_0^i[0], \ldots, \hat{u}_0^i[L'-1]\}$, where $L' \leq L$ is the list size at the decoding level $i$. The integer $i$ can be represented using the binary representation with $\nbf{i} = [i_0, i_1, \ldots, i_{n-1}]$. The binary representation of $i$ can be used to trace the sub-graphs at each stage which the $i$th bit channel belongs to.
Starting from the $(n-1)$th bit, if $i_{n-1} = 0$, the upper sub-graph is employed, else the lower sub-graph is employed at stage $m=1$. Similarly, given the sub-graph at stage $m=1$, we may determine the sub-graph at stage $m=2$ by the value of $i_{n-2}$.
For example, consider an $N=8$ polar code as given in Figure \ref{fig:dec8}. If $i=5$, then $\nbf{i} = [1, 0, 1]$. The last bit $i_2 = 1$ means that the lower sub-graph at stage $m=1$ and its corresponding input LLRs $\Gamma_1^{(4)}$, $\Gamma_1^{(5)}$, $\Gamma_1^{(6)}$, and $\Gamma_1^{(7)}$ are employed to decode the $i$th bit. Similarly, $i_1=0$ means that the subgraph with input LLRs $\Gamma_2^{(4)}$ and $\Gamma_2^{(5)}$ at stage $m=2$ are employed. Finally, $i_0=1$ means that $\Gamma_3^{(5)}$ is employed. There are $2^{n-m}$ LLR values required to store in the memory bank at stage $m$. For SCL decoding with shared memory, the number of memory blocks employed at stage $m$ depends on how many different patterns in ${\cal U}^{(i)}$ that results in different LLRs at stage $m$. Let $i(m)$ be the binary truncated version of $i$ which reserves only the $m$ most significant bits (MSBs) of $\nbf{i}$, i.e.,
\[
     i(m) = \sum_{j=n-m}^{n-1} i_j 2^j
\]
Define ${\cal U}_0^{(i(m)-1)}$ by
\[
    {\cal U}_0^{(i(m)-1)} = \{\hat{\nbf{u}}_0^{i(m)-1}: \forall \hat{\nbf{u}} \in {\cal U}^{(i)}\},
\]
and also define ${\cal U}_0^{(-1)} = \emptyset$. The set ${\cal U}_0^{(i(m)-1)}$ represents the subset obtained by reserving the first $i(m)$ bits of all candidates of ${\cal U}^{(i)}$.
Then it can be easily proved that, at the $i$th decoding level, the number of memory blocks to store the LLRs at stage $m$, denoted as $S_m(i)$, is given by
\beq
      S_m(i) = \min\left(|{\cal U}_0^{(i(m)-1)}|, 1 \right)
      \label{eq:sm}
\eeq
Based on the analysis of memory utilization at each decoding level, to limit the space complexity, we may restrict the maximum number of memory blocks individually for each stage. Note that the block size to store LLRs at stage $m$ is $2^{n-m}$. Therefore, it is more efficient to have a small number of memory blocks for small $m$. Define a vector $\nbf{L} = [L_1, \ldots, L_{n}]$, where $L_m$ is defined as the maximum number of memory blocks at stage $m$ with $L_m \leq L_{m'}$ for $m < m'$ and $L_n = L$. Then the total space complexity is in the order of
\beq
    \sum_{m = 1}^{n} L_m 2^{n-m}.
    \label{eq:s_comp}
\eeq
If $L_m = L$ for all $m=1, \ldots, n$, the decoder becomes a regular SCL decoder with list size $L$. Since the number of candidates at stage $m$ is reduced, the time complexity can also be reduced as well which is in the order of
\beq
    N\sum_{m=1}^n L_m,
\eeq
instead of $L N\log_2N=LNn$ for a regular SCL decoder of list size $L$.

Now the problem is how to restrict the maximum number of memory blocks for each stage. After the path extension at level $i$, we may obtain a set of candidates ${\cal U}^{(i)}$. 
For the SCL decoder, the best $L$ candidates among ${\cal U}^{(i)}$ are reserved which is independent of the decoding level $i$. However, to reduce the space complexity, our proposal reserves a variable number of candidates depending on the decoding level $i$. If $i$ locates at the sub-graph boundary at stage $m$, then the number of memory blocks required at stage $m$ for future extensions depends on how many candidates are reserved at level $i$.
Define $f(i)$ as an integer depending on $i$ which is the least integer of $j$ such that $i_j = 1$. 
For example, let $n=8$ and $i = 40$, then $\nbf{i}=[0, 0, 0, 1, 0, 1, 0, 0]$. The least integer of $j$ such that $i_j = 1$ is $j=3$ and therefore $f(40)=3$ for $n=8$.
It can be easily seen that the $i$th bit channel locates at the boundary of sub-graph at stage $m = n-f(i+1)$.
Therefore, to restrict the maximum number of memory blocks employed at each stage, we fist compute $m=n-f(i+1)$ and then reserve only $L_m$ candidates at decoding level $i$. In this way, the maximum number of memory blocks employed at the $m$th stage can be restricted to be $L_m$. Note that the case when $i = 2^n-1$ may require a special handling. In this case, we will $L_n$ candidates are reserved to left as many as possible candidates which may provides better performance with CRC-aided polar codes.

Table \ref{tab:iml} illustrates the numbers of reserved candidates at each decoding level $i$ to fulfil the memory constraint of $\nbf{L} = [4, 5, 6, 7]$ for   decoding of a polar code of length $N=16$. The low-complexity decoder has a certain performance loss as compared to the regular SCL decoder with $L=7$, because the number of candidates reserved at some decoding level is less than 7. However, the candidates in the list of SCL decoder get more reliable as $i$ increased. In this proposal, the smallest level $i$ that results in reserving $L_m$ candidates is $2^{n-m}-1$. Therefore, small $L_m$ values are applied only when $i$ is large which may results in a very small performance degradation. For example, given the smallest value $L_1=4$ as illustrated in Table \ref{tab:iml}, $L_1=4$ is applied only at level $i=7$.

In the next section, we proposed a multi-CRC encoding scheme for polar codes which helps to further reduce the number of memory blocks at certain decoding stages for the R-SCL decoder.

\begin{table}
\centering
\caption{The numbers of reserved candidates at each decoding level $i$ for $n=4$ with $\nbf{L}=[4, 5, 6, 7]$. \label{tab:iml}}
\begin{tabular}{| r l |} \hline
$i$:& \  0 \   1 \   2 \   3 \   4 \  5 \  6 \  7 \  8 \  9  10  11  12  13  14  15 \\ \hline
$m$:&    \ 4  \ 3 \   4 \   2 \   4 \  3 \  4 \  1 \  4 \  3  \  4 \  2  \  4 \ \ 3 \  4 \ \  4 \\ \hline
$L_m$:&   \ 7 \  6 \   7 \   5 \  7 \   6 \   7 \   4 \   7 \   6 \   7 \   5 \   7 \ \  6 \   7 \ \  7 \\ \hline
\end{tabular}
\end{table}

\section{Reduced-Complexity SCL Decoding of Multi-CRC-Aided Polar Codes}
\label{sec:mCRC}
The idea of multi-CRC encoding scheme for polar codes was first proposed in \cite{Guo2016}. The method proposed in \cite{Guo2016} is to divide the entire $K$-bit message block into $M$ sub-blocks. All sub-blocks have equal length of $K'=K/M$. Let the set of unfrozen bits be ${\cal A}$ with $|{\cal A}| = K + r$. 
Each message sub-block is processed by one CRC encoder with $r'$ parity check bits, where $r' = r/M$. Define $\nbf{a} = [a_0, a_1, \ldots, a_{K-1}]$. The Multi-CRC coding scheme proposed in \cite{Guo2016} partitions the $K$-bit message block into $M$ sub-blocks with the $j$th sub-block given by
\[
\nbf{a}_j = [a_{jK'}, \ldots, a_{(j+1)K'-1}], \ \ \ \ \mbox{for $j=0, \ldots,M-1$}.
\]
The multi-CRC encoding scheme results in the CRC encoded vector
\[
  \bar{\nbf{a}} = [\nbf{a}_0, \nbf{c}_0, \nbf{a}_1, \nbf{c}_1, \ldots, \nbf{a}_{M-1}, \nbf{c}_{M-1}],
\]
where $\nbf{c}_j$ is the $r'$-tuple CRC vector for the $j$th sub-block $\nbf{a}_j$. Then the multi-CRC encoded vector is mapped to the vector $\nbf{u}$ by letting $\nbf{u}_{\cal A} = \bar{\nbf{a}}$ and $\nbf{u}_{{\cal A}^c}=\nbf{0}$. Finally, the codeword is obtained by $\nbf{x} = \nbf{u} \nbf{G}_2^{\otimes n}$. In order to reduce the decoding delay, the modified SCL decoder proposed in \cite{Guo2016} outputs each sub-block as early as possible during the SCL decoding. When the decoding level of the modified SCL decoder reaches the last bit of the $j$th sub-block, the decoder applies CRC detection on the $j$th sub-block immediately among all candidates in the list. The most reliable candidate that passes the CRC check is selected as the estimation of the $j$th information sub-block. In this way, the latency can be reduced and the memory to store the {\it hard-decision} candidates can be released. However, the space complexity of the SCL decoder remains the same, because only the memory to store the {\it hard-decision} candidates are released. The memory size required to store the LLRs at each stage remains the same.

Our multi-CRC encoding scheme is different from that proposed in \cite{Guo2016}. The new design aims to reduce the memory size to store LLRs inside the R-SCL decoder which is more critical than to reduce the size of the hard-decision memory. The encoding method is described as follows. Let $M=2^s$ where $s\geq 1$.  Let the set of unfrozen bits be ${\cal A}$ with $|{\cal A}|=K+r$, where $K$ is the number of message bits and $r$ is the total number of CRC bits. Define ${\cal A}_j = {\cal A} \cap \{j2^{n-s}, \ldots, (j+1)2^{n-s}-1\}$ for $j=0, \ldots, 2^s-1$. Let $K_j=|{\cal A}_j|-r_j$ with $\sum_{j=0}^{2^s-1} K_j = K$ and $\sum_{j=0}^{2^s-1} r_j = r$, where $K_j$ is the number of message bits of the $j$th sub-block and $r_j$ is the number of CRC bits of the $j$th sub-block. Define the lengths of CRC bits as a vector $\nbf{r} = [r_0, r_1, \ldots, r_{2^s-1}]$. In the polar code design for multi-CRC encoding scheme, we require to select $|{\cal A}| = K + r$ unfrozen bit channels.
The message block $\nbf{a}$ is divided into $M$ sub-blocks with the size of the $j$th sub-block being $K_j$, i.e.,
\[
     \nbf{a} = [\nbf{a}_0, \nbf{a}_1, \ldots, \nbf{a}_{M-1}],
\]
where $\nbf{a}_j$ is a $K_j$-tuple vector. Note that the $K$-bit message block is partitioned into sub-blocks of different sizes which is quiet different from that proposed in \cite{Guo2016}. In addition, the numbers of CRC bits may be different for different sub-blocks. In our design, two types of CRCs are used. The first type is called ``local CRC'' which is added individually for each of the first $M-1$ sub-blocks. The second type is called ``global CRC'' which is added at the end of the block with $\nbf{a}$ as the message block. Define $\mbox{CRC}_j(\nbf{b})$ as the $j$th CRC encoding function with $\nbf{b}$ as the message block. The multi-CRC encoder results in the CRC encoded vector
\[
  \bar{\nbf{a}} = [\nbf{a}_0, \nbf{c}_0, \nbf{a}_1, \nbf{c}_1, \ldots, \nbf{a}_{M-1}, \nbf{c}_{M-1}],
\]
where $\nbf{c}_j = \mbox{CRC}_j(\nbf{a}_j)$, for $j=0, \ldots, M-2$, are the local CRC vectors and $\nbf{c}_{M-1}=\mbox{CRC}_{M-1}(\nbf{a})$ is the global CRC vector. The CRC encoded vector is then mapped to bit channels by $\nbf{u}_{\cal A} = \bar{\nbf{a}}$ and $\nbf{u}_{{\cal A}^c} = \nbf{0}$. Finally, the codeword is obtained by $\nbf{x} = \nbf{u} \nbf{G}_2^{\otimes n}$.

\begin{figure}[!t]
\centering
\includegraphics[width=\columnwidth]{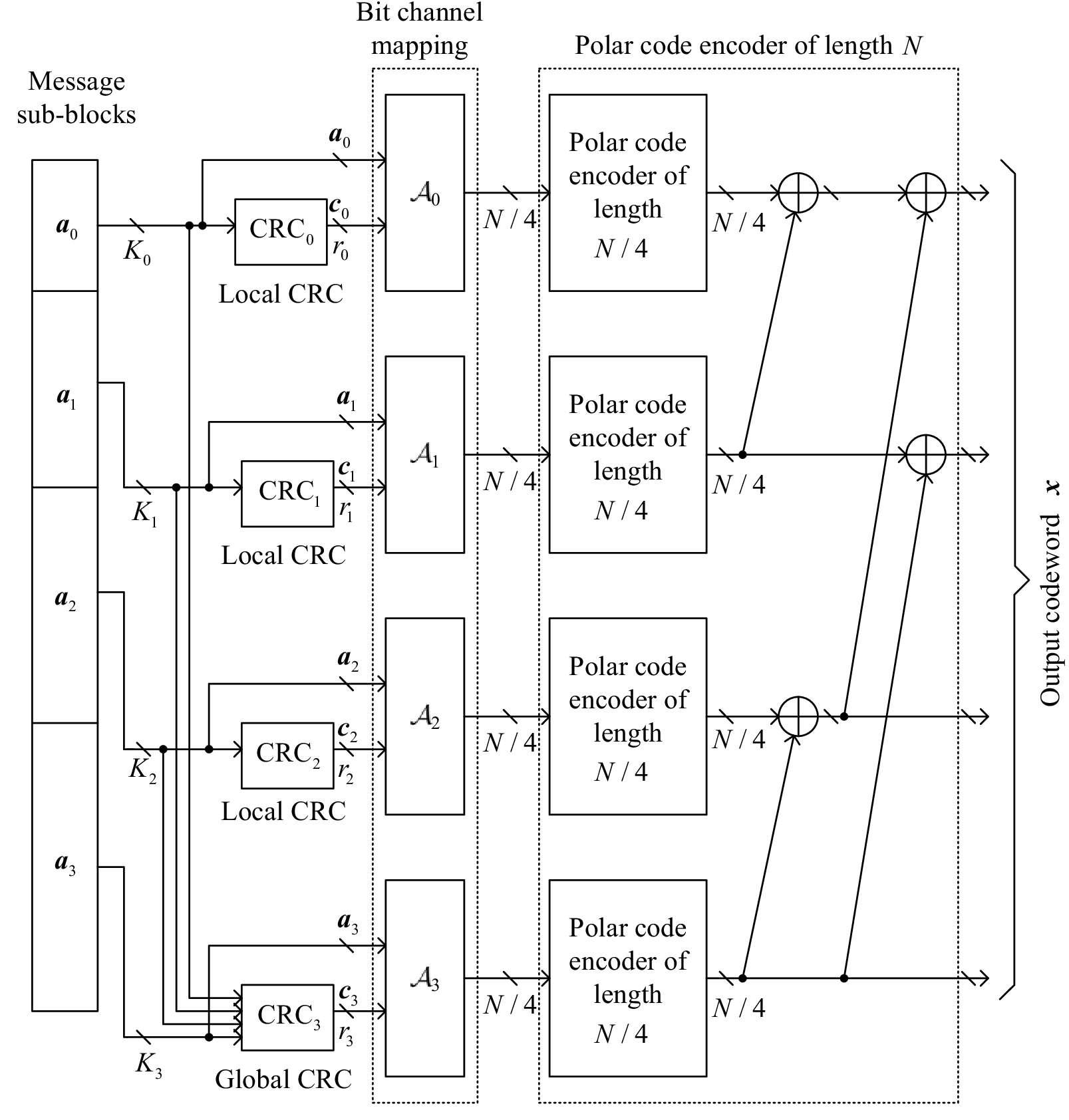}
\caption{Proposed multi-CRC encoder for polar codes of length $N$ with $M=4$ CRC sub-blocks.}
\label{fig:mCRC_enc}
\end{figure}

Figure \ref{fig:mCRC_enc} illustrates a block diagram of the encoder with $M=4$. This encoding method ensures that, after bit mapping, every consecutive $2^{n-s}$ bit channels contain exactly one message sub-block and one CRC vector. This property is very important for a modified R-SCL decoder to reduce the numbers of memory blocks to store LLRs from stage $1$ to stage $s$. Given the setting of $\nbf{L}=[L_1, \ldots, L_n]$ as that proposed in Section \ref{sec:R-SCL}. The modified R-SCL decoder for the proposed multi-CRC-aided polar code performs the same as that proposed in Section \ref{sec:R-SCL}, except when the decoding level $i$ reaches the  end of every $2^{n-s}$ bit channels, i.e., when $i = (j+1)2^{n-s}-1$ for some integer $j \geq 0$.
Whenever $i = (j+1)2^{n-s}-1$ for some integer $j\geq 0$, the candidates for the $j$th message sub-block and its corresponding CRC vector are ready in the list. The modified R-SCL decoder performs local CRC check on each of the candidates for the $j$th sub-block and selects the $L_{s-f(j+1)}$ best candidates that pass the $\mbox{CRC}_j$ check for $j=0, \ldots, M-2$. If $i$ reaches the last decoding level, i.e., $i=2^n-1$, the decoder performs global CRC check and selects only one best candidate that passes the global $\mbox{CRC}_{M-1}$ check. The local CRC detections exclude a large number of candidates that do not pass the CRC check. As can be seen in Section \ref{sec:R-SCL}, a large size of memory to store the LLRs from stage $1$ to stage $s$ can be saved by setting small values of $L_m$ for $m \leq s$.

One important issue for the new proposal is to determine how many CRC bits should be employed for the local CRC and global CRC. A larger number of local CRC bits provides more reliable detection of each sub-block for the modified R-SCL decoder and hence we can set a small value of $L_m$ for $m \leq s$. However, using too many local CRC bits also degrades the performance even with a large list size. The reason is that, given a fixed $K$, more unreliable bit channels are used by increasing the number of local CRC bits, and hence the probability that the correct codeword is not in the final candidates is also increased.
 Therefore, we should carefully select the numbers of local CRC bits that just allow to set small values of $L_m$ for $m \leq s$ without much performance impact.

\section{Simulation Results}
\label{sec:simu}
\subsection{Reduced-Complexity SCL Decoding of a Single-CRC-Aided Polar Code}
\begin{figure}[!t]
\centering
\includegraphics[width=\columnwidth]{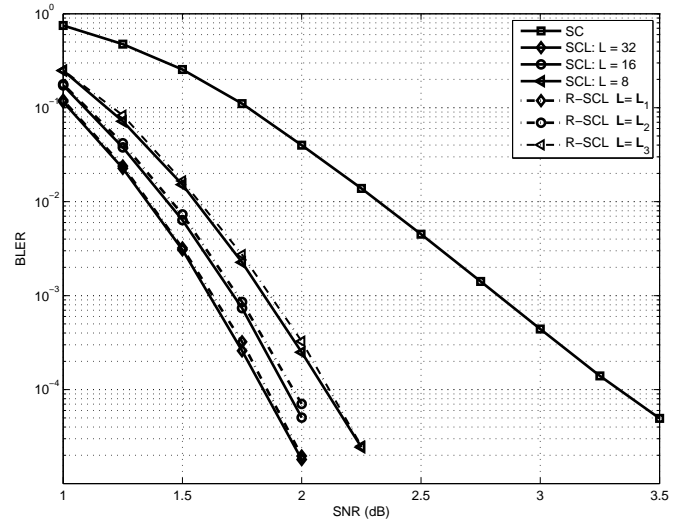}
\caption{Performance of R-SCL decoding of a single-CRC-aided polar code.}
\label{fig:single-crc}
\end{figure}

A single CRC-aided $(2048, 1024)$ polar code with 16 CRC bits is employed in the simulation with the CRC polynomial $x^{16}+x^{12}+x^5+1$. Three different configurations of R-SCL decoders are considered with the following list size vectors
\bean
     \nbf{L}_1 &=& [22,    24,    26,    28,    30,    32,    32,    32,    32,    32,    32],\\
     \nbf{L}_2 &=& [11,    12,    13,    14,    15,    16,    16,    16,    16,    16,    16],\\
     \nbf{L}_3 &=& [5,     6,     7,     7,     7,     8,     8,     8,     8,    8,    8].
\eean
The space and time complexities of the configurations are given in Table \ref{tab:comp1}, as well as those of regular SCL decoders with $L=32$, $L=16$, and $L=8$. It should be noted that the R-SCL decoders with $\nbf{L} = \nbf{L}_1$, $\nbf{L}=\nbf{L}_2$, and $\nbf{L}=\nbf{L}_3$ are considered as the R-SCL versions of the SCL decoders with $L=32$, $L = 16$, and $L = 8$, respectively. The space and time complexities of the R-SCL decoders are reduced by about 25\% and 8\%, respectively. Simulation results of the block error rates (BLERs) of all the above configurations are shown in Figure \ref{fig:single-crc}. The results indicate that the single CRC-aided R-SCL decoders have slight performance losses which are at most 0.03dB at high SNR regions as compared to those of SCL decoders. The space complexity can be further reduced by using the multi-CRC-aided polar codes as shown in the next subsection.

\begin{table}
\centering
\caption{The space and time complexities of SCL and R-SCL decoders.\label{tab:comp1}}
\begin{tabular} {|l|l|l|} \hline
Decoder & Space Complexity & Time Complexity \\ \hline \hline
SCL: $L=32$ & 65504 & 720896 \\ \hline
R-SCL $\nbf{L} = \nbf{L}_1$ & 48992 (74.8\%) & 659456 (91.5\%) \\ \hline \hline
SCL: $L=16$ & 32752 & 360448 \\ \hline
R-SCL $\nbf{L} = \nbf{L}_2$ & 24496 (74.8\%) & 329728 (91.5\%)\\ \hline \hline
SCL: $L=8$ & 16376 & 180224 \\ \hline
R-SCL $\nbf{L} = \nbf{L}_3$ & 11832 (72.2\%)& 163840 (90.9\%)\\ \hline
\end{tabular}
\end{table}
\subsection{Reduced-Complexity SCL Decoding of Multi-CRC-Aided Polar Codes}
\begin{figure}[!t]
\centering
\includegraphics[width=\columnwidth]{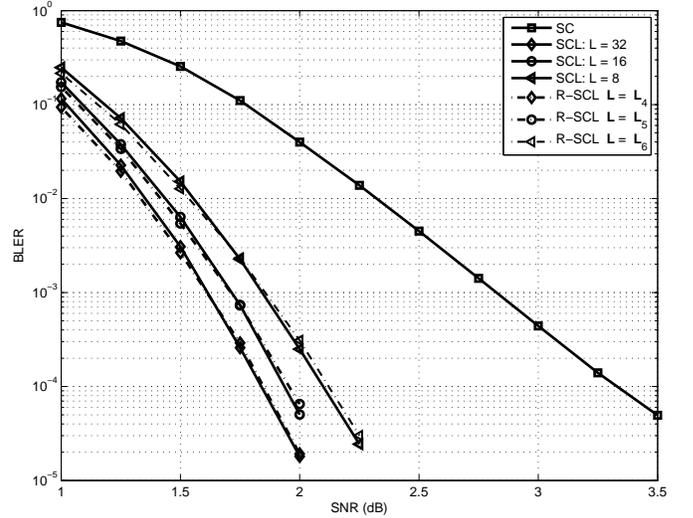}
\caption{Performance of the R-SCL decoding of a multi-CRC-aided polar code with $\nbf{r} = [2,2,2,10]$.}
\label{fig:multi-crc}
\end{figure}

We consider to partition the 1024-bit message block into $M = 4$ sub-blocks in the multi-CRC-aided polar code. The numbers of local CRC bits are all set to 2 and the number of global CRC bits is set to 10, i.e.,  $\nbf{r} = [2,2,2,10]$. The CRC polynomials of CRC-2 and CRC-10 are given by $x^2+x+1$ and $x^{10}+x^9+x^8+x^7+x^6+x^4+x^3+1$, respectively. Three different configurations of R-SCL decoders are considered with the following list size vectors
\bean
     \nbf{L}_4 &=& [8,    16,    32,    32,    32,    32,    32,    32,    32,    32,    32],\\
     \nbf{L}_5 &=& [4,    8,    16,    16,    16,    16,    16,    16,    16,    16,    16],\\
     \nbf{L}_6 &=& [2,     4,     8,     8,     8,     8,     8,     8,     8,    8,    8].
\eean
The space and time complexities of the configurations are given in Table \ref{tab:comp2}. The R-SCL decoders with $\nbf{L} = \nbf{L}_4$, $\nbf{L}=\nbf{L}_5$, and $\nbf{L}=\nbf{L}_6$ are considered as the R-SCL versions of the SCL decoders with $L=32$, $L = 16$, and $L = 8$, respectively. The space and time complexities of all the R-SCL decoders are reduced by about 50\% and 11\%, respectively. Simulated BLERs of all the above configurations are shown in Figure \ref{fig:multi-crc}. The results show that, at low SNR regions, the R-SCL decoders with the multi-CRC-aided polar code outperform the SCL decoders with the single-CRC-aided polar code. The reason is that R-SCL decoders with the multi-CRC-aided polar code may include more legal (local CRC passed) candidates at decoding levels of $i=(j+1)2^{n-s}-1$ for $j=0, \ldots, M-2$. This fact increases the probability that the correct codeword is in the final candidates.  At high SNR regions, the R-SCL decoders with the multi-CRC-aided polar code have slight losses which are at most 0.02dB compared to those of SCL decoders with the single-CRC-aided polar code.

\begin{table}
\centering
\caption{The space and time complexities of R-SCL decoders.\label{tab:comp2}}
\begin{tabular} {|l|l|l|} \hline
Decoder & Space Complexity & Time Complexity \\ \hline
R-SCL $\nbf{L} = \nbf{L}_4$ & 32736 (50\%) & 638976 (88.7\%)\\ \hline
R-SCL $\nbf{L} = \nbf{L}_5$ & 16368 (50\%) & 319488 (88.7\%)\\ \hline
R-SCL $\nbf{L} = \nbf{L}_6$ & 8184 (50\%)  & 159744 (88.7\%)\\ \hline
\end{tabular}
\end{table}

\begin{figure}[!t]
\centering
\includegraphics[width=\columnwidth]{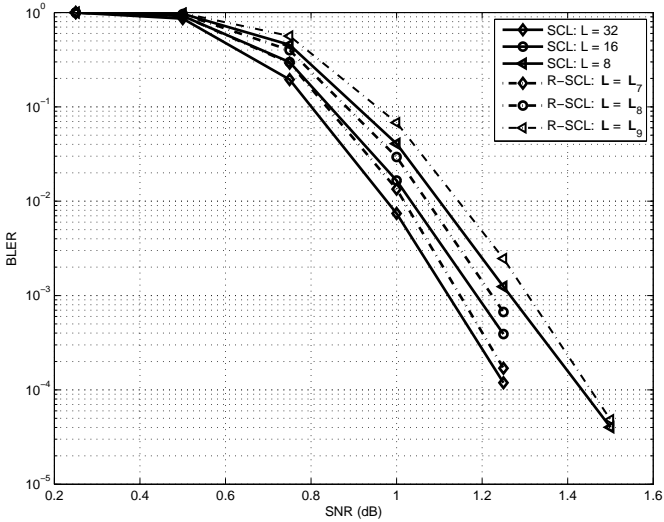}
\caption{Performance of the R-SCL decoding of a multi-CRC-aided $(16384,8192)$ polar code with $\nbf{r} = [10, 10, 10, 10, 10, 10, 10, 10]$.}
\label{fig:multi-crc3}
\end{figure}

A major concern of polar codes is that the space complexity of the SCL decoder becomes prohibitive for large code block sizes. In this simulation, we consider a multi-CRC-aided $(16384, 8192)$ polar code with $M=8$ and $\nbf{r} = [10,10,10,10, 10,10,10,10]$. The list size vectors considered are given by
\bean
     \nbf{L}_7 &=& [1,    1,   1,  32,    32,    32,    32,    32,    32,    32,    32,    32,  32,  32],\\
     \nbf{L}_8 &=& [1,    1,   1,  16,    16,    16,    16,    16,    16,    16,    16,    16,  16,  16],\\
     \nbf{L}_9 &=& [1,    1,   1,   8,     8,     8,     8,     8,     8,     8,    8,     8,    8,   8].
\eean
Since $L_1=L_2=L_3 = 1$ for all the configurations, these configurations correspond to hybrid SCL/SC decoders. The polar code of length 16384 is decoded using a SCL decoder of length 2048 for each sub-block of length 2048, and SC decoding scheme is applied in the sub-block level.
The space and time complexities of the configurations are given in Table \ref{tab:comp3}. The R-SCL decoders with $\nbf{L} = \nbf{L}_7$, $\nbf{L}=\nbf{L}_8$, and $\nbf{L}=\nbf{L}_9$ are considered as the R-SCL versions of the SCL decoders with $L=32$, $L = 16$, and $L = 8$, respectively. The space complexities of the R-SCL decoders are reduced by about 76\% to 85\% and the time complexities are reduced by about 18\% to 20\%.  Simulated BLERs shown in Figure \ref{fig:multi-crc3} indicate a worst performance loss of 0.04dB compared to those of SCL decoders with the single-CRC-aided polar code.

\begin{figure}[!t]
\centering
\includegraphics[width=\columnwidth]{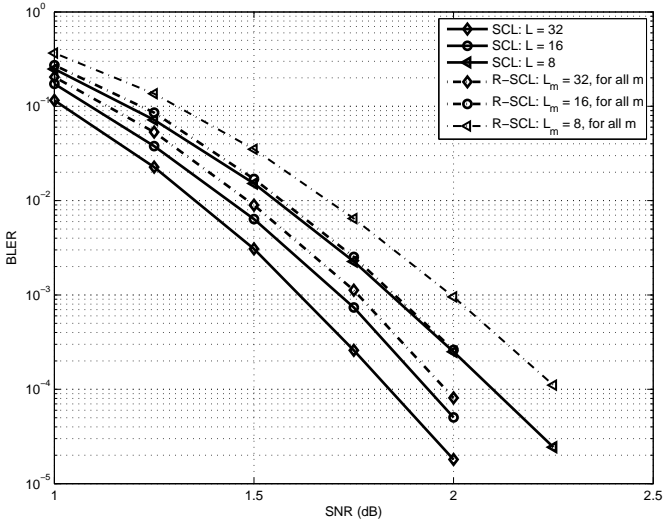}
\caption{Performance of the R-SCL decoding of a multi-CRC-aided $(2048,1024)$ polar code with $\nbf{r} = [10, 10, 10, 10]$.}
\label{fig:multi-crc4}
\end{figure}

We have conducted another simulation for a multi-CRC-aided $(2048, 1024)$ polar code with $M=4$ and $\nbf{r} = [10,10,10,10]$. The simulated BLERs are shown in Figure \ref{fig:multi-crc4} with full-complexity R-SCL decoders, i.e. $L_m=L$ for all $m$. The results indicate that even using the full-complexity decoders, the performance losses are very significant due to the heavy overhead of a total of 40 CRC bits for the short code. Therefore, for multi-CRC-aided polar codes, we suggest to use a small number of total CRC bits for short codes and a large number of total CRC bits for long codes.

\begin{table}
\centering
\caption{The space and time complexities of R-SCL decoders.\label{tab:comp3}}
\begin{tabular} {|l|l|l|} \hline
Decoder & Space Complexity & Time Complexity \\ \hline
R-SCL $\nbf{L} = \nbf{L}_7$ & 79840 (15.2\%) & 5816320 (79.2\%)\\ \hline
R-SCL $\nbf{L} = \nbf{L}_8$ & 47088 (17.9\%) & 2932736 (79.9\%)\\ \hline
R-SCL $\nbf{L} = \nbf{L}_9$ & 30712 (23.4\%)  & 1490944 (81.2\%)\\ \hline
\end{tabular}
\end{table}

\section{Conclusion}
\label{sec:conc}
This paper proposes a reduced-complexity SCL (R-SCL) decoding algorithm for polar codes. The R-SCL decoders are effective to reduced the space complexity while maintain acceptable performance levels. A design example of the $(2048, 1024)$ single-CRC-aided polar code shows that R-SCL decoders with 25\% reduction of space complexity and 8\% reduction of time complexity can still achieve almost the same performance levels as those decoded by SCL decoders. Multi-CRC-aided polar codes are proposed to further reduce the space complexities of R-SCL decoders. A design example of the $(2048, 1024)$ multi-CRC-aided polar code shows that R-SCL decoders with 50\% reduction of space complexity and 11\% reduction of time complexity have a worst performance loss of 0.02dB compared to a single-CRC-aided polar code decoded by SCL decoders. Another aggressive setting of the R-SCL decoder for the $(16384, 8192)$ multi-CRC-aided polar code with about 85\% reduction of space complexity and 20\% reduction of time complexity results in a worst performance loss of 0.04dB. Finally, we demonstrate an example showing how important to properly select the number of CRC bits for each sub-block in the design of multi-CRC-aided polar codes. We observed that adding too many CRC bits increases the probability that the correct codeword is not in the final candidates and hence degrades performance.


%

%

\section*{Acknowledgment}
The authors would like to thank W.-J.\ Chen and W.-N.\ Sun for helpful discussions.

\ifCLASSOPTIONcaptionsoff
  \newpage
\fi

\bibliographystyle{IEEEtran}


\end{document}